\documentclass[transmag,12pt]{IEEEtran}
\usepackage[shortlabels]{enumitem}
\usepackage{graphicx} 
\usepackage{epsfig} 
\usepackage{amsmath,amsthm,amssymb}  
\usepackage{float}
\usepackage{cases,setspace,adjustbox}
\usepackage{cite}
\usepackage{array}
\usepackage[update,prepend]{epstopdf}
\usepackage{eqparbox}
\usepackage{url}
\usepackage{caption, subcaption}
\usepackage{xcolor}
\usepackage{mathtools}
\usepackage{bbm}
\usepackage{multirow}
\usepackage{multicol}

\usepackage{booktabs}

\usepackage[acronym]{glossaries}
\usepackage{comment}
\newcommand\blfootnote[1]{%
	\begingroup
	\renewcommand\thefootnote{}\footnote{#1}%
	\addtocounter{footnote}{-1}%
	\endgroup
}
\setacronymstyle{long-short}

\newacronym{AR}{AR}{Augmented Reality}
\newacronym{ai}{AI}{Artificial Intelligence}
\newacronym{arima}{ARIMA}{Autoregressive Integrated Moving Average}

\newacronym{cdf}{CDF}{cumulative distribution function}
\newacronym{cnn}{CNN}{Convolutional Neural Network}
\newacronym{comp}{CoMP}{Coordinated Multi-Point}
\newacronym{convlstm}{ConvLSTM}{Convolutional LSTM}
\newacronym{cp}{CP}{Cyclic Prefix}
\newacronym{crs}{CRS}{Cell Specific
Reference Signal}
\newacronym{ctl}{CTL}{Communications Technology Laboratory}
\newacronym{CU}{CU}{Centralized Unit}

\newacronym{d2d}{D2D}{Device-to-Device}
\newacronym{dc}{DC}{dual connectivity}
\newacronym{dci}{DCI}{Downlink Control Information}
\newacronym{dl}{DL}{downlink}
\newacronym{dsrc}{DSRC}{Dedicated short-range communications}
\newacronym{dss}{DSS}{Dynamic Spectrum Sharing}
\newacronym{DU}{DU}{Distributed Unit}

\newacronym{ets}{ETS}{Exponential Smoothing}
\newacronym{embb}{eMBB}{enhanced mobile broadband}
\newacronym{enb}{eNB}{Evolved Node-B}
\newacronym{epc}{EPC}{Evolved Packet Core}
\newacronym{emtc}{eMTC}{enhanced MTC}

\newacronym{fdd}{FDD}{Frequency Domain Duplex}
\newacronym{fdm}{FDM}{Frequency Domain Multiplexing}
\newacronym{firstnet}{FirstNet}{First Responder Network Authority}

\newacronym{gan}{GAN}{Generative Adversarial Network}
\newacronym{gps}{GPS}{Global Positioning System}
\newacronym{gnb}{gNB}{Next Generation Node-B}
\newacronym{gbps}{Gb/s}{gigabits per second}

\newacronym{harq}{HARQ}{Hybrid Automatic Repeat reQuest}
\newacronym{haps}{HAPS}{High-Altitude Platform Stations}

\newacronym{iot}{IoT}{Internet of Things}
\newacronym{iiot}{IIoT}{Industrial Internet of Things}
\newacronym{iomt}{IoMT}{Internet of Medical Things}
\newacronym{irs}{IRS}{Intelligent Reflecting Surfaces}
\newacronym{its}{ITS}{Intelligent Transport Systems}
\newacronym{itu}{ITU}{International Telecom Union}

\newacronym{kpi}{KPIs}{Key Performance Indicators}

\newacronym{lte}{LTE}{Long Term Evolution}
\newacronym{lteapro}{LTE-A~Pro}{LTE Advanced Pro}
\newacronym{lstm}{LSTM}{Long Short-Term Memory}

\newacronym{mcs}{MCS}{Modulation and Coding Scheme}
\newacronym{mcptt}{MCPTT}{Mission-Critical Push-to-Talk}
\newacronym{mimo}{MIMO}{Multiple-Input Multiple-Output}
\newacronym{mmtc}{mMTC}{massive machine-type communication}
\newacronym{mtc}{MTC}{Machine-Type Communication}
\newacronym{mlp}{MLP}{Multilayer Perceptron}
\newacronym{ma}{MA}{Moving Average}
\newacronym{mm}{MM}{Moving Median}
\newacronym{ml}{ML}{Machine Learning}

\newacronym{nist}{NIST}{National Institute of Standards and Technology}
\newacronym{nbiot}{NB-IoT}{Narrowband-Internet of Things}
\newacronym{nr}{NR}{New Radio}
\newacronym{ntia}{NTIA}{National Telecommunications and Information Administration}
\newacronym{ntn}{NTN}{Non-Terrestrial Networks}

\newacronym{oran}{O-RAN}{Open Radio Access Network}
\newacronym{owl}{OWL}{Online Watcher of LTE}

\newacronym{prosas}{ProSAS}{Proactive Spectrum Adaptation Scheme}
\newacronym{prb}{PRBs}{Physical Resource Blocks}
\newacronym{psc}{PSC}{Public Safety Communications}
\newacronym{pscch}{PSCCH}{Physical Sidelink Control Channel}
\newacronym{psfch}{PSFCH}{Physical Sidelink Feedback Channel}
\newacronym{pssch}{PSSCH}{Physical Sidelink Shared Channel}

\newacronym{qos}{QoS}{Quality of Service}

\newacronym{ran}{RAN}{Radio Access Network}
\newacronym{rans}{RANs}{Radio Access Networks}
\newacronym{rat}{RAT}{Radio Access Technology}
\newacronym{re}{RE}{Resource Element}
\newacronym{ric}{RIC}{RAN Intelligent Controller}
\newacronym{relu}{ReLU}{Rectified Linear Unit}
\newacronym{rmse}{RMSE}{root-mean-square error}
\newacronym{rnn}{RNN}{Recurrent Neural Network}
\newacronym{rl}{RL}{Reinforcement Learning}
\newacronym{RU}{RU}{Radio Unit}
\newacronym{rt}{RT}{real-time}

\newacronym{sdr}{SDR}{Software Defined Radio}
\newacronym{sl}{SL}{sidelink}
\newacronym{smo}{SMO}{Service and Management Orchestration}

\newacronym{td3}{TD3}{Twin Delayed Deep Deterministic Policy Gradient}
\newacronym{tdd}{TDD}{Time Domain Duplex}
\newacronym{tdm}{TDM}{Time Domain Multiplexing}
\newacronym{timegan}{TimeGAN}{Time-Series Generative Adversarial Network}
\newacronym{tr}{TR}{Technical Report}
\newacronym{thz}{THz}{terahertz}
\newacronym{TIP}{TIP}{Telecom Infra Project}

\newacronym{urllc}{URLLC}{ultra-reliable low-latency communication}
\newacronym{uav}{UAVs}{Unmanned Aerial Vehicles}
\newacronym{ue}{UE}{User Equipment}
\newacronym{ul}{UL}{uplink}

\newacronym{v2v}{V2V}{Vehicle-to-Vehicle}
\newacronym{v2x}{V2X}{Vehicle-to-Everything}
\newacronym{VR}{VR}{Vitual Reality}
\newacronym{vnf}{VNFs}{Virtualized Network Functions}
\newacronym{vm}{VMs}{Virtual Machines}

\newacronym{wnd}{WND}{Wireless Networks Division}
\newacronym{wlan}{WLAN}{Wireless Local Area Network}
\newacronym{xr}{XR}{Extended Reality}
\newacronym{3gpp}{3GPP}{3rd Generation Partnership Project}
\usepackage{soul}
\begin{document}
\title{O-RAN and 6G: The Future of Wireless Innovation?}
\author{Sneihil Gopal$^{\ddagger\dagger}$, Aziz Kord$^{\star}$, and Richard A. Rouil$^{\star}$\\
$^{\ddagger}$PREP Associate, National Institute of Standards and Technology (NIST), USA\\
$^{\dagger}$Department of Physics, Georgetown University, USA\\
$^{\star}$National Institute of Standards and Technology (NIST), USA.\\
Emails: \{sneihil.gopal, azizollah.kord, richard.rouil\}@nist.gov}
\maketitle
\section*{Abstract}
\label{sec:abstract}
The emergence of 6G technology represents a significant advancement in wireless communications, providing unprecedented speed, extremely low latency, and pioneering applications. In light of this development, an important question arises: Can the Open Radio Access Network (O-RAN), with its emphasis on openness, flexibility, RAN slicing, RAN Intelligent Controller (RIC), and cost-effectiveness, fulfill the complex requirements of 6G? This paper delves into the potential synergy between O-RAN and 6G, illustrating how O-RAN can facilitate customization, reduce expenses, and stimulate innovation in next-generation networks. We also tackle the challenges associated with 6G, such as the need for exceptional performance, integration with non-terrestrial networks, and heightened security. By examining the interaction between O-RAN and 6G, we underscore their joint role in shaping the future of wireless communication. Lastly, we demonstrate O-RAN's potential through a unique, learning-based spectrum-sharing solution that aligns with 6G's objectives for efficient spectrum usage.
\section{Introduction}
\label{sec:intro}
\gls{oran}~\cite{ref:ORAN_polese_2023}, an industry-driven initiative, represents a significant change in wireless network architecture. It embraces the principles of openness, flexibility, intelligence, and cost-efficiency, fundamentally transforming the design and deployment of \gls{ran}. This is achieved by replacing proprietary, monolithic solutions with disaggregated, interoperable components. The open and collaborative framework not only reduces barriers to entry for smaller vendors but also empowers network operators to customize and optimize their networks to meet specific needs. As a result, \gls{oran} has gained significant traction within the telecommunications industry, providing both greenfield and brownfield service providers with a pathway to operate more agilely, optimize costs, and innovate at a faster pace. By enabling multi-vendor ecosystems, \gls{oran} enhances network flexibility and performance, making it a critical driver in the evolution of next-generation wireless communication. \blfootnote{U.S. Government work, not subject to U.S. Copyright.}

Simultaneously, the development of 6G represents a significant leap forward in wireless communication, aiming to expand on the groundwork laid by 5G. It promises enhanced speed, ultra-low latency, and the capacity to accommodate a wide array of transformative applications~\cite{ref:6G_IMT2030}. Going beyond just enhancing the capabilities of 5G, 6G foresees a future where wireless networks can facilitate advanced technologies ranging from \gls{xr} to cutting-edge healthcare solutions and smart cities~\cite{ref:6G_de_2021_survey}. To achieve data rates surpassing 100~\gls{gbps}, reducing latency to mere fractions of a millisecond, and providing connectivity for millions of devices per square kilometer, 6G is set to revolutionize digital interaction, pushing the boundaries of connectivity and redefining the Information Age.

Given the rapid technological progress, a significant question arises: Can \gls{oran}, with its core principles of openness, flexibility, intelligence, and cost-effectiveness, seamlessly align with the diverse and demanding requirements of 6G? This article starts by introducing the fundamental concepts of \gls{oran} and 6G, creating a context for their roles in the wireless communication landscape. It offers a detailed analysis of how these technologies can work together, highlighting the benefits of their integration, such as improved network agility, cost-effectiveness, and greater innovation potential. To provide a practical perspective, the article includes a novel \gls{rl}-based \gls{dss} solution within the \gls{oran} framework. We discuss the deployment of the intelligent spectrum-sharing solution within the \gls{oran} architecture along with the end-to-end workflow. The solution illustrates how \gls{oran} can enhance spectrum efficiency, aligning with the objectives of 6G. It also discusses the challenges and considerations of combining these technologies, including important issues like interoperability, security, and regulatory compliance. Through this thorough exploration, the article aims to explain how \gls{oran} and 6G can collectively shape the future of wireless communication, creating a more connected, efficient, and adaptable digital ecosystem.
\section{Background}
\label{ref:background}
In the fast-evolving world of wireless communication, it's crucial to understand the fundamentals of \gls{oran} and the goals of 6G. This section offers a comprehensive overview of both \gls{oran} and 6G, laying the groundwork for their integration.
\subsection{O-RAN Architecture and Principles}
\gls{oran} represents a major shift in the design, deployment, and operation of wireless networks. Unlike traditional, proprietary \gls{ran} architectures, \gls{oran} emphasizes openness, flexibility, intelligence, and cost-efficiency. By disaggregating the \gls{ran} components and promoting interoperability among vendors, \gls{oran} fosters innovation, lowers costs, and allows network operators to customize their networks according to specific requirements. This section explores the architecture and core principles of \gls{oran}, highlighting its transformative potential for the wireless industry.

\begin{figure}
    \centering
    \includegraphics[width=\columnwidth]{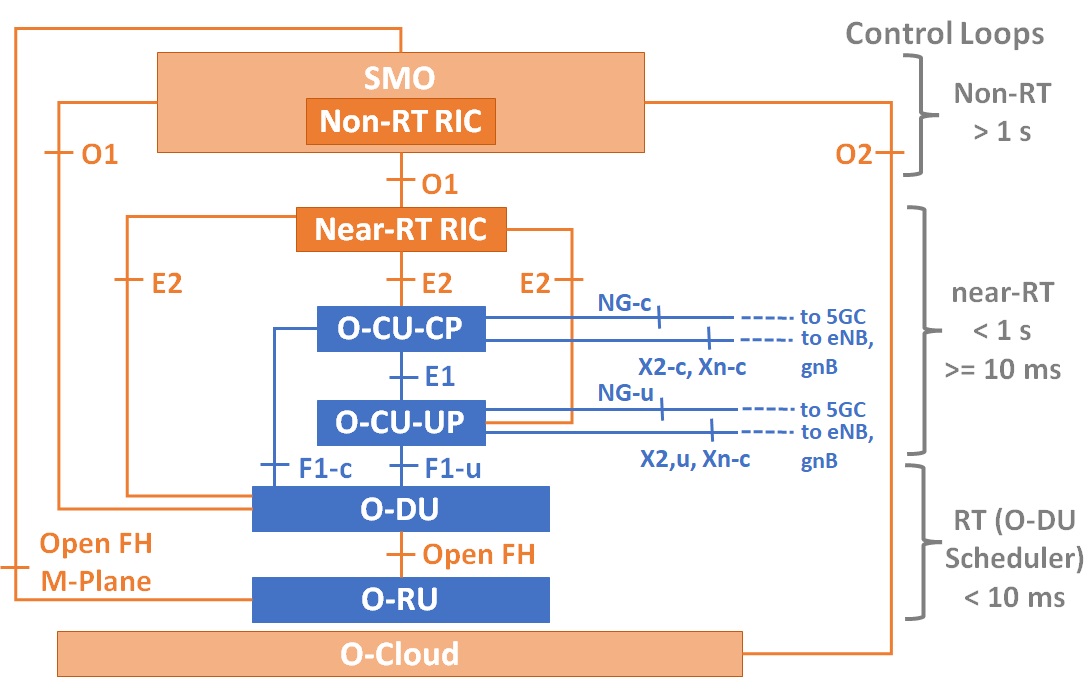}
    \caption{Overall O-RAN Architecture. The ``blue'' elements and interfaces are specified by the \gls{3gpp}, whereas ``orange'' elements and interfaces are specified by \gls{oran} Alliance~\cite{ref:oran_tr_wg1}.}
    \label{fig:architecture}
     \vspace{-1.25em}
\end{figure}

\textbf{Disaggregated RAN Architecture:} At the core of \gls{oran} is its disaggregated architecture, which divides the traditional monolithic \gls{ran} into smaller, modular components. Fig.~\ref{fig:architecture} illustrates the overall \gls{oran} architecture 
which includes the \gls{RU}, \gls{DU}, \gls{CU}, and \gls{ric}~\cite{ref:ORAN_polese_2023}. 

The \gls{RU} in the \gls{oran} architecture is used to transmit and receive radio signals over the air interface and handle lower-layer processing, such as modulation, demodulation, and beamforming. Different \gls{oran} configurations allow operators to deploy \gls{RU}s tailored to specific use cases or geographic regions, enhancing network efficiency and flexibility. The \gls{DU} manages \gls{rt} processing tasks like radio link control, medium access control, and portions of the physical layer, acting as the link between the \gls{RU} and the \gls{CU}. This disaggregation enables the deployment of processing capabilities closer to the network edge, reducing latency and improving performance for applications that require rapid response times. The \gls{CU} is responsible for non-\gls{rt} processing, which includes higher-layer protocols, control plane functions, and mobility management. It is usually deployed in centralized data centers and uses cloud computing resources for scalability and flexibility. Its separation from the \gls{DU} allows for greater network flexibility and independent scaling of components.

The \gls{ric} is a significant innovation in \gls{oran}. It provides centralized control and optimization of \gls{ran} functions using \gls{ai} and \gls{ml} algorithms. The \gls{ric} is divided into two layers: the near-\gls{rt} \gls{ric}, which manages tasks requiring rapid responses such as resource management and interference mitigation, and the non-\gls{rt} \gls{ric}, which focuses on long-term tasks like network planning and optimization. This layered approach allows for immediate and strategic network improvements, enhancing overall \gls{ran} performance.

\textbf{Openness and Interoperability:} \gls{oran} is committed to openness and interoperability. Unlike traditional \gls{ran} systems that rely on proprietary solutions from a single vendor, \gls{oran} is designed to be vendor-neutral by using standardized interfaces and protocols. This allows different \gls{ran} components to communicate seamlessly, regardless of the vendor. The approach specifies open interfaces within \gls{oran}, such as the fronthaul interface (Open FH) between the \gls{RU} and \gls{DU}, the midhaul interface (F1-u and F1-c) between the \gls{DU} and \gls{CU}, and the interfaces for the \gls{ric} (O1 and E2), shown in Fig.~\ref{fig:architecture}, enabling control and management functions. This standardization empowers operators to mix and match components from different vendors, promoting a competitive and innovative \gls{ran} ecosystem.

To ensure interoperability between vendors, the \gls{TIP} has introduced a badging system with gold, silver, and bronze tiers. This system fosters a competitive market for \gls{oran} solutions, enabling operators to choose the best components for their needs. These efforts promote openness, interoperability, and the development of a more adaptable and resilient communication infrastructure~\cite{ref:oran_TIP}.


\textbf{Virtualization and Cloud-Native Design:} \gls{oran} leverages virtualization and cloud-native design principles to greatly enhance the flexibility, scalability, and efficiency of the \gls{ran}. By separating software functions from hardware, \gls{oran} allows network functions to be deployed on standard, off-the-shelf hardware, reducing costs and simplifying network operations. In the \gls{oran} framework, traditional \gls{ran} functions are implemented as \gls{vnf}. These \gls{vnf} can be deployed on \gls{vm} or containers running on standard x86 servers~\cite{ref:oran_arnaz2022toward}. This approach enables network operators to allocate resources dynamically based on demand, scale network functions as required and deploy these functions closer to the network edge to support low-latency applications.

In addition to virtualization, \gls{oran} adopts a cloud-native architecture, designing and deploying network functions as microservices capable of running in containerized environments. Cloud-native principles such as automation, orchestration, and continuous integration/deployment empower operators to achieve greater agility in their networks, automate various management tasks, and rapidly deploy new services and features. This combination of virtualization and cloud-native approaches allows \gls{oran} to offer a highly flexible and efficient solution for modern RAN deployments.

\textbf{AI/ML-Driven Optimization:} \gls{ai}/\gls{ml} plays a central role in \gls{oran}, especially within the framework of the \gls{ric}. By integrating \gls{ai}/\gls{ml} algorithms into the \gls{ric}, \gls{oran} enables intelligent, data-driven optimization of the \gls{ran}. This capability is essential for managing the complexity of modern wireless networks and meeting the diverse demands of different applications and users.

The near-\gls{rt} \gls{ric} utilizes \gls{ai}/\gls{ml} algorithms to optimize \gls{ran} functions that require rapid responses, such as dynamic spectrum allocation, load balancing, and interference management. These optimizations are driven by real-time network data, allowing the \gls{ran} to adapt to changing conditions and enhance overall performance swiftly. On the other hand, the non-\gls{rt} \gls{ric} is dedicated to long-term network optimization, leveraging \gls{ai}/\gls{ml} for tasks such as traffic prediction, network planning, and energy efficiency improvements. The non-\gls{rt} \gls{ric} can make informed decisions that improve network performance while reducing operational costs by analyzing historical data and learning from network behavior over time. This dual-layered approach ensures that \gls{oran} can continuously adapt and optimize itself in the short and long term.

\textbf{Cost-Efficiency and Innovation:} The combination of openness, interoperability, virtualization, and \gls{ai}-driven optimization in \gls{oran} brings about significant cost efficiencies and fosters innovation within the wireless industry. By promoting a competitive, multi-vendor ecosystem, \gls{oran} reduces reliance on proprietary solutions, thereby lowering both 
capital and operational expenditures for network operators. In addition, the disaggregation of \gls{ran} components, along with the adoption of open interfaces and standard hardware, allows operators to avoid vendor lock-in and reduce the costs associated with deploying and maintaining their networks. Lastly, virtualization plays a key role in these cost savings by enabling more efficient use of hardware resources and minimizing the need for specialized equipment.

In summary, \gls{oran}’s open and flexible architecture empowers operators to innovate and customize their networks to meet specific needs. Whether it's deploying new services, optimizing network performance for particular applications, or integrating \gls{ai}-driven solutions, \gls{oran} provides the necessary tools and framework for continuous innovation. This environment not only enhances network performance but also allows operators to stay ahead in a rapidly evolving technological landscape.

\subsection{6G: Vision and Requirements}
\begin{figure}
    \centering
    \includegraphics[width=\columnwidth]{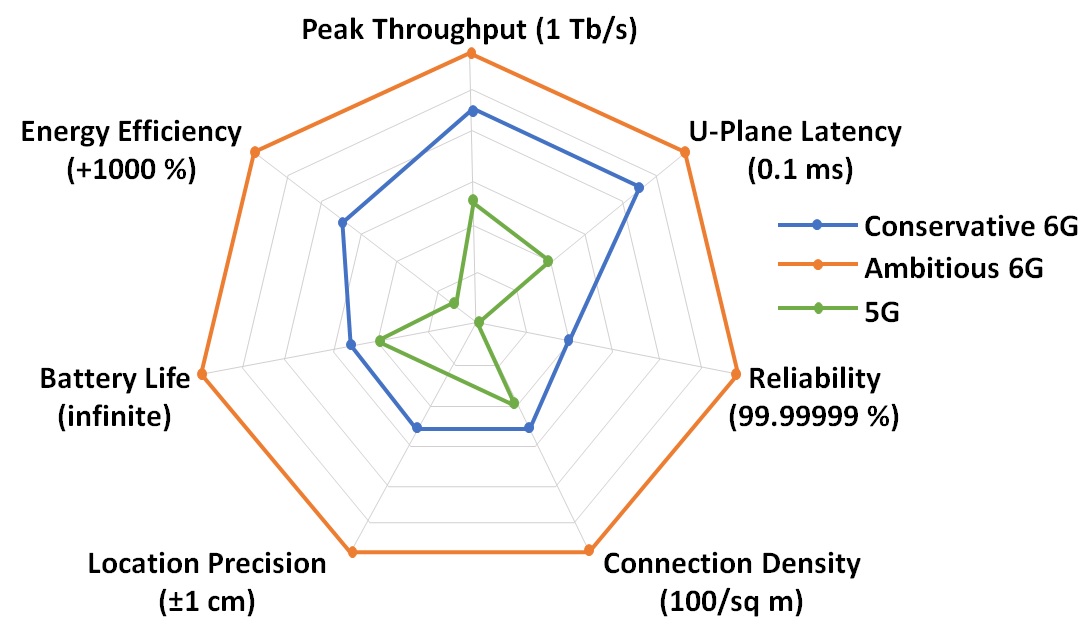}
    \caption{Comparison of 5G and 6G \gls{kpi}.}
    \label{fig:6g}
     \vspace{-1.25em}
\end{figure}

The introduction of 6G represents a major leap in wireless communications, building on the advancements of 5G. While 5G significantly improved speed, latency, and connectivity, 6G is expected to further elevate these capabilities, ushering in a new era of digital interaction with unprecedented features and transformative applications~\cite{ref:6G_saad_2019_vision}. Fig.~\ref{fig:6g} compares 5G and 6G, highlighting both conservative and ambitious targets for 6G, which demonstrate significant improvements over 5G~\cite{ref:6G_Keysight}. In this section, we explore the key objectives and requirements shaping future networks, offering insights into the ambitious goals of 6G.

\textbf{Ultra-High Data Rates and Capacity:} One of the most significant advancements promised by 6G is the achievement of ultra-high data rates, expected to surpass 100 \gls{gbps}. This leap in speed will facilitate the seamless transmission of vast amounts of data, supporting increasingly data-intensive applications such as immersive \gls{xr}, 8K video streaming, and real-time telepresence. Additionally, 6G aims to dramatically increase network capacity, allowing for the simultaneous connection of millions of devices per square kilometer. This is particularly critical in urban environments and smart cities, where dense populations and the proliferation of \gls{iot} devices demand robust and scalable network solutions.

\textbf{Ultra-Low Latency:} Latency is another crucial aspect of 6G development. While 5G has already made significant strides in reducing latency to around 1 millisecond, 6G aspires to reduce it even further, to the realm of microseconds. This ultra-low latency is essential for applications that require real-time responsiveness, such as autonomous vehicles, industrial automation, remote surgery, and immersive \gls{xr}. The ability to transmit data almost instantaneously will enable new levels of precision and reliability in critical applications, driving innovation across multiple industries.

\textbf{Integration of \gls{ntn}:} 6G is envisioned as a truly global network, integrating terrestrial and non-terrestrial elements to provide seamless coverage across the entire planet. This includes the incorporation of satellite communication systems, \gls{haps}, and \gls{uav} to deliver connectivity in remote and underserved regions. The integration of \gls{ntn} will not only enhance global coverage but also improve network resilience and reliability, ensuring uninterrupted service even in the most challenging environments. This aspect of 6G is particularly relevant for disaster recovery, remote monitoring, and global logistics.

\textbf{Advanced Spectrum Utilization:} The exponential growth in connected devices and data traffic necessitates more efficient use of the 
spectrum. 6G aims to address this challenge through advanced spectrum-sharing techniques and the utilization of higher frequency bands, including \gls{thz} frequencies. These higher frequencies offer greater bandwidth and 
extremely high data rates, but they also present challenges related to signal propagation and attenuation. To overcome these challenges, 6G will rely on sophisticated techniques such as beamforming, massive \gls{mimo}, and \gls{irs}. Additionally, dynamic spectrum sharing strategies, powered by \gls{ai}/\gls{ml}, will be critical in optimizing spectrum efficiency and maximizing network performance.

\textbf{Enhanced Security and Privacy:} As 6G expands the scope and scale of wireless networks, it also introduces new security and privacy challenges. The increased reliance on \gls{ai}, the integration of \gls{ntn}, and the proliferation of \gls{iot} devices create a more complex and potentially vulnerable network environment. To address these concerns, 6G will need to incorporate advanced security mechanisms, including quantum-resistant cryptography, \gls{ai}-driven threat detection, and decentralized security architectures. Privacy-preserving technologies, such as differential privacy and secure multi-party computation, will play a crucial role in safeguarding user data and ensuring trust in 6G networks.

\textbf{\gls{ai}-Driven Network Optimization:} \gls{ai}/\gls{ml} will be at the core of 6G network management and optimization. These technologies will enable real-time decision-making, predictive analytics, and autonomous network operation, driving efficiency and performance to new heights. \gls{ai} will be instrumental in managing the complexity of 6G networks, from dynamic spectrum allocation to resource optimization and fault detection. The integration of \gls{ai} will also facilitate the customization of network services, allowing operators to tailor performance parameters to specific use cases and user requirements.

\begin{figure}
    \centering
    \includegraphics[width=0.95\columnwidth]{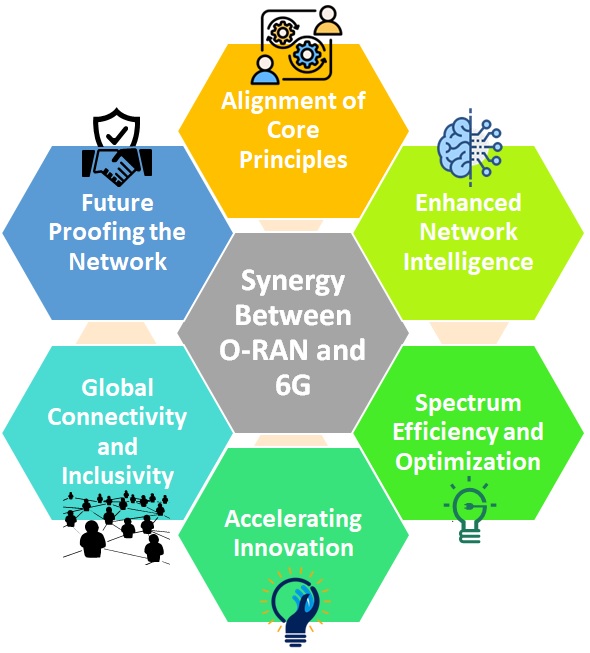}
    \caption{Synergy between \gls{oran} and 6G.}
    \label{fig:synergy}
     \vspace{-1.25em}
\end{figure}

\textbf{Sustainability and Energy Efficiency:} As the demand for connectivity grows, so does the need for sustainable and energy-efficient network solutions. 6G is expected to prioritize green technologies and energy-saving techniques, reducing the carbon footprint of wireless networks. This includes the development of energy-efficient hardware, use of renewable energy sources, and implementation of \gls{ai}-driven energy management systems. The focus on sustainability is not only a response to environmental concerns but also a necessity for the long-term viability of expansive and dense 6G networks.

While \gls{oran} and 6G are on parallel trajectories, each with its own set of principles, objectives, and technological advancements, the convergence of these two innovations represents a strategic alignment, where \gls{oran}'s openness, flexibility, and agility can seamlessly integrate with 6G's ambitious goals. In the following sections, we explore why \gls{oran} and 6G are a perfect match and the transformative potential of this union.

\section{Synergy Between \gls{oran} and 6G}
\label{sec:perfect_match}
The integration of \gls{oran} with 6G technology is not just a convergence of two distinct innovations, but a synergistic relationship. \gls{oran} provides 6G with the best path to achieve its objectives and has the potential to redefine wireless communication~\cite{ref:ORAN_6G}. This section explores the multifaceted synergy between \gls{oran} and 6G, highlighting how their combined strengths can accelerate the evolution of future networks.

\textbf{Alignment of Core Principles:} 
\gls{oran}’s openness and flexibility naturally complement 6G’s vision of ultra-high-performance, adaptable networks, enabling more efficient and scalable connectivity. By embracing open interfaces and disaggregated components, \gls{oran} allows 6G networks to be more adaptable and responsive to the diverse requirements of future applications. This alignment is crucial for enabling the scalability and customization needed to support the vast array of use cases anticipated in the 6G era, from \gls{xr} to smart cities and beyond.

\textbf{Enhanced Network Intelligence:} The connection between \gls{oran} and 6G is particularly evident in the realm of network intelligence. \gls{oran}’s \gls{ric}, which integrates \gls{ai}/\gls{ml} algorithms, is a critical enabler for the advanced, \gls{ai}-driven optimizations that 6G demands. The near-\gls{rt} and non-\gls{rt} \gls{ric} allows for dynamic resource management, predictive maintenance, and automated decision-making processes, all of which are essential for maintaining the \gls{urllc} and \gls{mmtc} that 6G networks aim to provide. This intelligence not only enhances network efficiency but also ensures that 6G networks can adapt to the dynamic needs of users and applications in real time.

\textbf{Spectrum Efficiency and Optimization:} Spectrum management is a crucial challenge for 6G networks. These networks operate across a wide range of frequencies, including sub-THz bands. \gls{oran}’s open architecture allows for more efficient spectrum sharing and dynamic spectrum allocation, which are vital for optimizing the use of this limited resource. By utilizing \gls{ai}-driven spectrum management solutions within the \gls{oran} framework, 6G networks can achieve higher levels of spectrum efficiency. This enables faster data rates, reduced interference, and overall better network performance. This collaboration is vital as 6G aims to support unprecedented data demands and provide seamless connectivity in diverse environments.

\textbf{Accelerating Innovation:} The open, modular nature of \gls{oran} promote a competitive ecosystem of vendors and solution providers, which is instrumental in driving innovation in the 6G landscape. By lowering barriers to entry and enabling greater experimentation and collaboration, \gls{oran} creates an environment where new ideas and technologies can rapidly be tested and deployed. This openness is vital for 6G, which will require constant innovation to meet the evolving demands of future applications. Whether it’s developing new \gls{ai} algorithms, creating more efficient hardware, or pioneering novel use cases, the synergy between \gls{oran} and 6G accelerates the pace of innovation, ensuring that the network remains at the forefront of technological advancement.

\textbf{Global Connectivity and Inclusivity:} One of the most significant capabilities between \gls{oran} and 6G is the potential to democratize access to advanced wireless technologies. \gls{oran}’s emphasis on openness and interoperability can help reduce costs and increase the availability of 6G networks in underserved and remote regions. By enabling a more diverse ecosystem of vendors and service providers, \gls{oran} ensures that 6G can be deployed more widely, helping to bridge the digital divide and promote global connectivity. This synergy is not only about technical alignment but also about the shared goal of creating a more inclusive and connected world.

\textbf{Future-Proofing the Network:} As 6G continues to advance, the flexibility and adaptability of \gls{oran} will be crucial in future-proofing the network. The modular architecture of \gls{oran} allows for continuous upgrades and the integration of new technologies, ensuring that 6G networks can evolve without requiring complete overhauls. This capability is essential for maintaining the relevance and performance of 6G networks over the long term, as new applications and challenges emerge.

The synergy between \gls{oran} and 6G, as illustrated in Fig.~\ref{fig:synergy}, not only strengthens the capabilities of both technologies but also paves the way for a new era of connectivity that is faster, more reliable, intelligent, and inclusive. By aligning their core principles, enhancing network intelligence, optimizing spectrum usage, accelerating innovation, and promoting global connectivity, \gls{oran} and 6G collaboratively create a robust and adaptable network infrastructure capable of 
supporting the next generation of wireless communications.
\section{\gls{rl}-Based \gls{dss} Solution}
\label{sec:spectrum_sharing}
With growing wireless communication demand, efficient spectrum utilization is critical. \gls{oran} and 6G technologies offer intelligent \gls{ai}/\gls{ml}-based spectrum management solutions. In this section, we introduce an \gls{rl}-based \gls{dss} solution specifically designed for \gls{oran} environments. The proposed solution leverages \gls{rl} for intent-driven spectrum management, focusing on minimizing resource surpluses and deficits in \gls{ran}s~\cite{ref:oran_gopal_2024_adapshare}. Spectrum management techniques often rely on static allocation, which can lead to inefficiencies like over-provisioning or under-provisioning of resources~\cite{ref:oran_gopal_2024prosas}. In contrast, this \gls{rl}-driven approach dynamically adapts to real-time network conditions, making it highly suitable for the complex and variable nature of 6G networks. 
The \gls{rl} agent uses the \gls{td3} algorithm to dynamically adapt spectrum allocation based on real-time network conditions, learning optimal policies through feedback~\cite{ref:rl_sutton2018reinforcement}.

A key aspect of this \gls{rl}-based approach is its dynamic adaptation to network conditions, which sets it apart from traditional static allocation methods. The \gls{rl} agent is designed to find an optimal policy by following a ``state, action, reward, next-state'' trajectory. This optimal policy maximizes the cumulative discounted reward, expressed as: $R_t = \sum_{t=0}^{T} \gamma^{t} r_{t},\quad 0 \le \gamma \le 1$, where $\gamma$ is the discount factor and $T$ is the time horizon~\cite{ref:rl_sutton2018reinforcement}.

\begin{figure}
    \centering
    \includegraphics[width=\columnwidth]{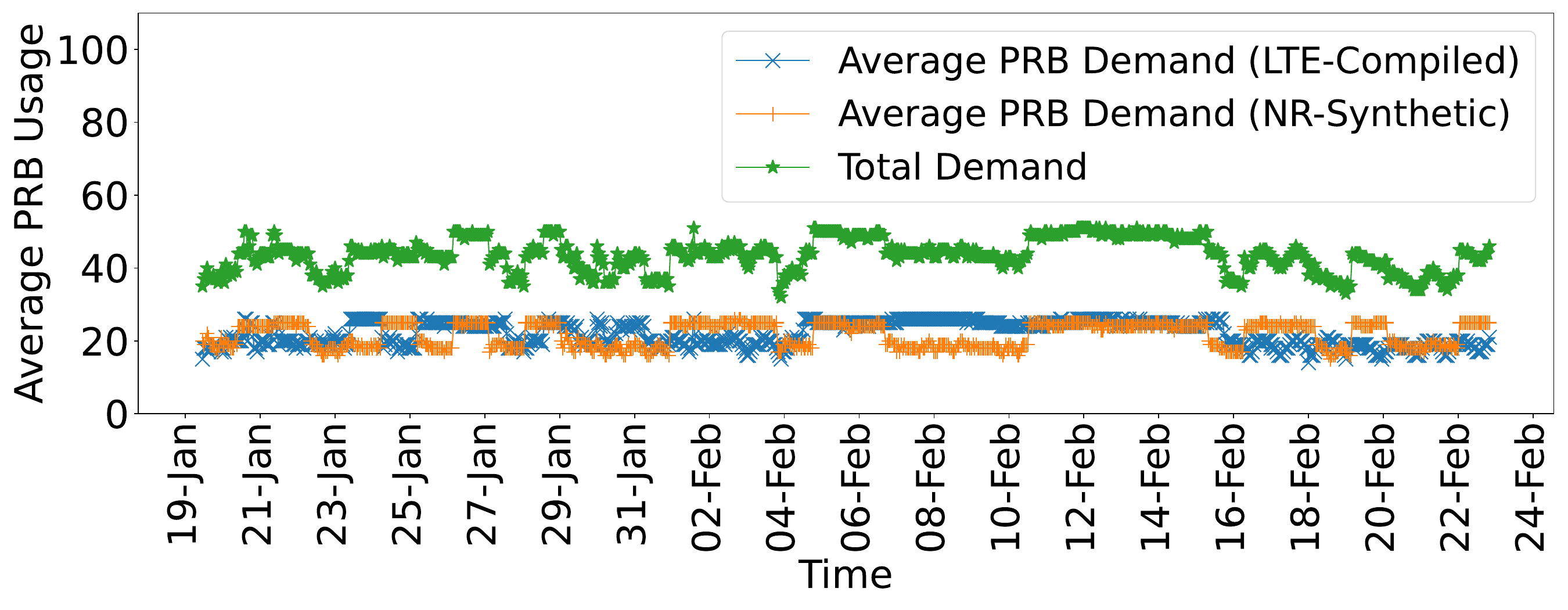}
    \caption{Average \gls{prb} usage/demand vs. Time.}
    \label{fig:dataset}
    \vspace{-1.25em}
\end{figure}

The resource allocation problem is formulated as a contextual bandit problem, which allows the \gls{rl} agent to make decisions based on the context provided by both current and historical network data. The agent observes the resource demands, i.e., the number of \gls{prb} required, of the coexisting \gls{rans} at various time steps. The state space $S$ includes ordered pairs of these demands at the current and previous time steps. For example, at time $t$, the agent observes resource demand at $t$ and $n$ previous time steps, denoted as $o_{t} = \{(D_{A,t},D_{B,t}),\dots,(D_{A,t-n},D_{B,t-n})\}$, where $(D_{A,k}$ and $D_{B,k})$ represent the resource demand of $\text{RAN}_\text{A}$ and $\text{RAN}_\text{B}$, respectively, at time $k$. The action space $A$ consists of all possible partitions of the resource pool of $N_r$ \gls{prb} between the two networks. Each action $a_t \in A$ corresponds to a specific allocation of $N_{A,t}$ and $N_{B,t}$ \gls{prb} to $\text{RAN}_\text{A}$ and $\text{RAN}_\text{B}$, respectively, at time $t$, where the actions are continuous-valued. The goal of the \gls{rl} agent is to maximize the current reward, which is defined as $r_t = - J - \zeta J$, where $J$ is the weighted sum of squared fractional surpluses or deficits in the \gls{ran}s and is defined as: $J = \zeta\left(\frac{N_{A,t}-D_{A,t}}{D_{A,t}}\right)^2 + (1-\zeta) \left(\frac{N_{B,t}-D_{B,t}}{D_{B,t}}\right)^2$, where, $\zeta \in [0,1]$ is the weighting factor that enables intent-driven spectrum management by allowing the controller to prioritize one network over the other. For instance, $\zeta \rightarrow 1$ gives a higher priority to $\text{RAN}_\text{A}$, while $\zeta \rightarrow 0$ prioritizes $\text{RAN}_\text{B}$.

The \gls{rl} agent was trained using a comprehensive dataset of \gls{lte} scheduling information collected at NIST’s Gaithersburg campus between January and February 2023~\cite{ref:datset_adapshare}. This dataset, captured from downlink traffic at 2115 MHz in Band 4 using OWL~\cite{ref:owl_bui_2016}, an online
decoder of the LTE control channel, 
includes detailed LTE \gls{dci} such as system frame number (SFN), subframe index, radio network temporary identifiers (RNTIs), the number of \gls{prb} allocated to devices, modulation and coding scheme (MCS), and DCI message type. These data were transformed into a time-series representation of PRB allocations for $\text{RAN}_\text{A}$. 

Additionally, a synthetic dataset for $\text{RAN}_\text{B}$ PRB allocations was generated using \gls{timegan}, a tool that creates synthetic time-series data while preserving temporal dependencies~\cite{ref:timegan}. This synthetic data mirrors the statistical properties of the real-world \gls{lte} data, under the assumption that the coexisting network share \gls{lte}-compatible numerology with 15~kHz subcarrier spacing and an identical time-frequency resource grid, serving similar devices as \gls{lte}. Fig.~\ref{fig:dataset} shows the average PRB demand for $\text{RAN}_\text{A}$ and $\text{RAN}_\text{B}$ over time. This time-series highlights the dynamic resource demand of networks and provides insights into the spectrum usage patterns.

\begin{figure}
    \centering
    \includegraphics[width=\columnwidth]{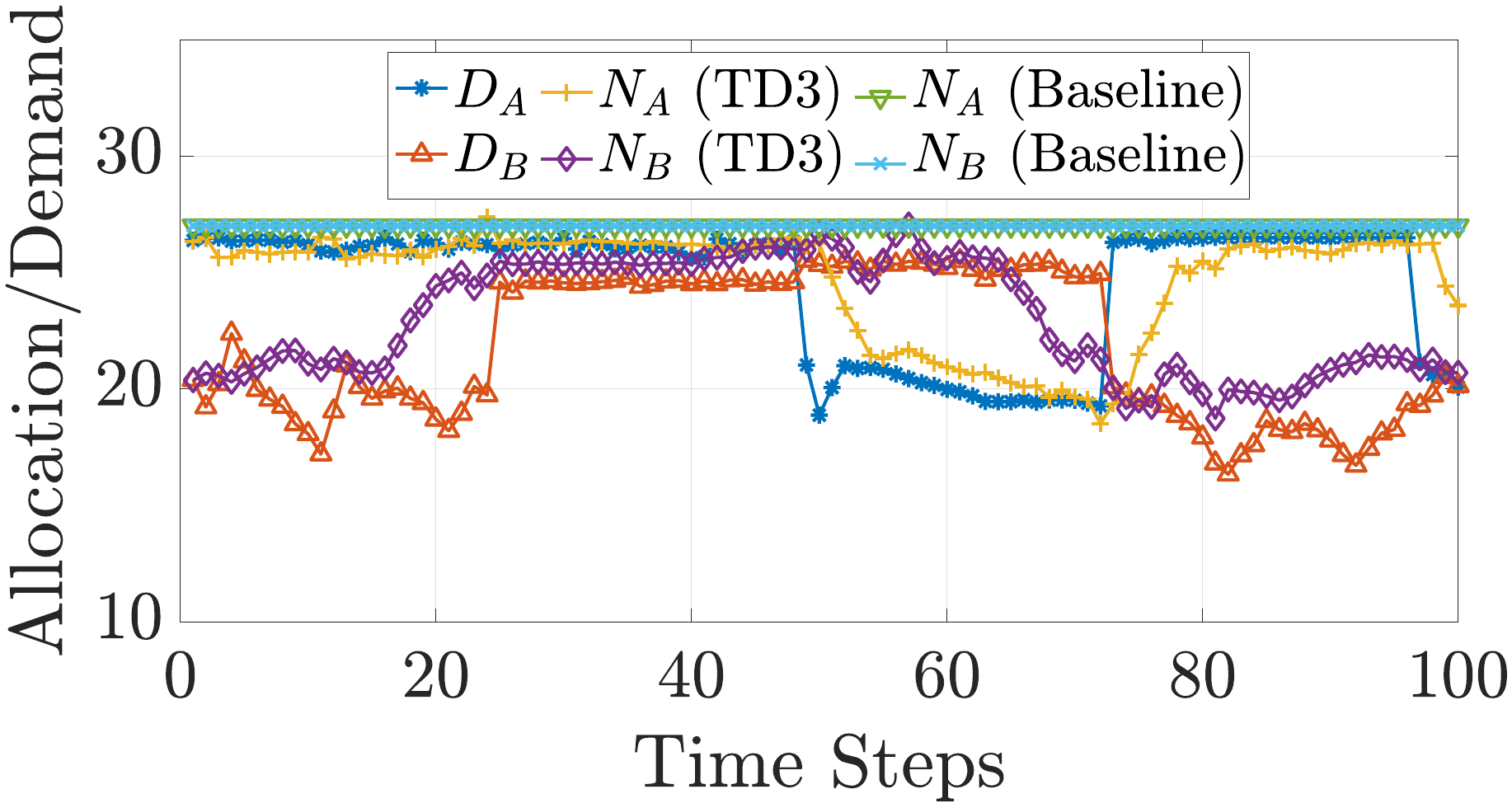}
    \caption{Average \gls{prb} demand/allocation vs. Time for $\zeta = 0.5$.}
    \label{fig:demand_alloc}
     \vspace{-1.25em}
\end{figure}

Fig.~\ref{fig:demand_alloc} presents a comparison of PRB demand and allocation when the proposed \gls{rl}-based solution is applied, versus a quasi-static approach that allocates resources based on peak demand. The \gls{rl}-based approach significantly improves efficiency by allocating resources dynamically based on real-time demand, whereas the static approach leads to over-provisioning of resources. The graph illustrates how the proposed method results in more balanced and efficient spectrum usage, reducing waste and improving network performance.

\begin{figure*}
    \centering
    \includegraphics[width=\textwidth]{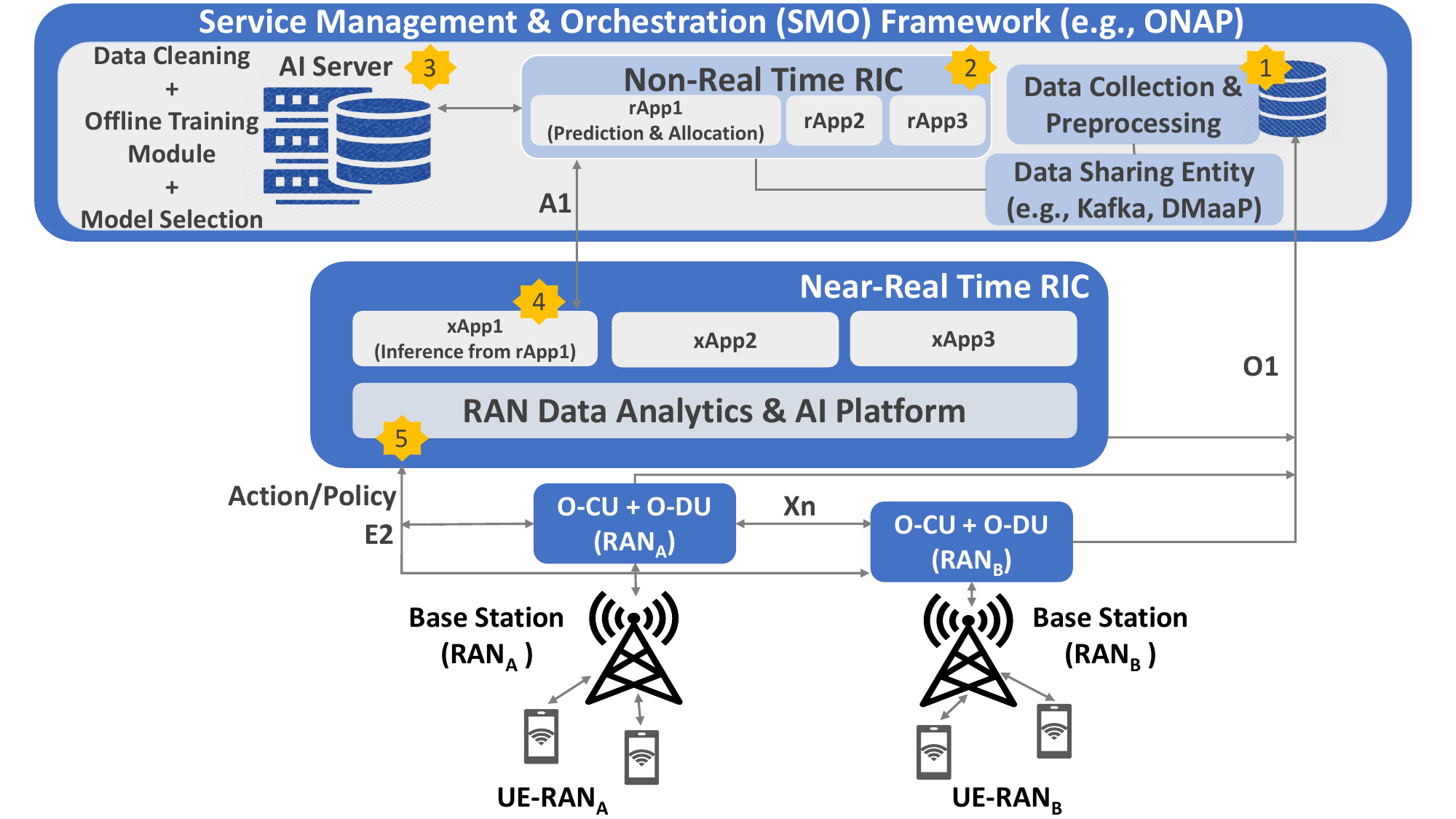}
    \caption{High-level structure illustrating the deployment of the intelligent spectrum-sharing solution as an rApp within the O-RAN architecture~\cite{ref:oran_tr_wg1}.}
    \label{fig:deployment}
\end{figure*}

The proposed solution can be flexibly deployed as either an rApp in the non-\gls{rt} \gls{ric} or as an xApp in the near-\gls{rt} \gls{ric}. When deployed as an rApp, the solution operates within the non-\gls{rt} \gls{ric}, focusing on long-term data analysis and optimization. The policies generated by the rApp are communicated to the near-\gls{rt} \gls{ric} for real-time implementation, making this deployment ideal for applications that do not have stringent latency requirements. Fig.~\ref{fig:deployment} illustrates the overall deployment of the proposed solution within the \gls{oran} architecture as an rApp. Alternatively, in the near-\gls{rt} \gls{ric}, the \gls{rl} agent operates as an xApp, allowing for real-time resource allocation adjustments in response to immediate network demands. This scenario is particularly suited for fast-changing network environments where low latency and quick decision-making are critical. 

By integrating this \gls{rl}-based solution into the \gls{oran} framework, network operators can achieve several key benefits. The system’s ability to dynamically adapt to near real-time changes in network conditions ensures that spectrum resources are used efficiently, even in highly variable environments. Additionally, the continuous learning process helps to reduce interference and improve overall \gls{qos}. The solution is also scalable, making it well-suited for managing the increasing number of connected devices anticipated in 6G networks. In summary, the \gls{rl}-based \gls{dss} solution aligns seamlessly with the goals of \gls{oran} and 6G, particularly in enhancing spectrum efficiency and enabling more dynamic network management. By optimizing spectrum utilization, this solution paves the way for a more intelligent, adaptable, and efficient wireless communication landscape.
\section{Challenges and Future Directions}
Integrating \gls{oran} with 6G presents several technical, operational, and regulatory challenges. Overcoming these challenges is essential to fully realizing the potential of this integration and driving the future of wireless communication. 

\textbf{Interoperability and Standardization:} 
Interoperability remains a key challenge due to varying vendor implementations, making rigorous testing and standardized protocols essential for seamless integration. Future efforts should focus on developing universal standards and protocols that accommodate diverse vendor implementations, critical for achieving an open and interoperable network environment.

\textbf{Ultra-Low Latency Requirements:} 
While \gls{oran}’s disaggregated architecture offers flexibility and cost-efficiency, it introduces potential latency at component interfaces. To meet 6G’s stringent demands, optimizing these interfaces and enhancing real-time processing capabilities will be crucial. Future research should prioritize refining network architectures and developing new technologies that minimize latency, supporting \gls{urllc} necessary for advanced applications.
    
\textbf{Security Concerns:} 
\gls{oran}’s openness introduces security vulnerabilities. Critical challenges include protecting the network from cyber threats, ensuring data integrity, and maintaining robust access control in a multi-vendor environment. Future directions include developing advanced security frameworks and integrating \gls{ai}-driven security solutions to safeguard 6G deployments within \gls{oran} architecture. Research should also explore robust encryption methods and anomaly detection systems tailored to \gls{oran}’s open architecture, enhancing the overall security of these networks.
    
\textbf{Complexity of Network Management:} 
\gls{oran}’s flexibility increases the complexity of network management in 6G, requiring advanced automation tools. \gls{ai}/\gls{ml} will play vital roles in simplifying network management and predicting potential issues before they impact performance. Future efforts should develop more sophisticated \gls{ai}/\gls{ml} algorithms to automate network optimization and maintenance tasks, ensuring efficient and reliable network operations.
    
\textbf{Scalability:} Scalability is another critical challenge as networks grow in size and complexity. Ensuring that \gls{oran} solutions can scale effectively involves the technical scalability of \gls{ran} components and the efficient management and optimization of large-scale deployments. Future research should focus on creating scalable architectures and management frameworks capable of handling the expansive demands of 6G, with modular and flexible network designs vital to achieving this scalability.

\textbf{Spectrum Allocation and Management:} Integrating \gls{oran} with 6G necessitates careful consideration of spectrum allocation and management policies. \gls{dss} is essential for maximizing spectrum efficiency, but it must be supported by regulatory frameworks that allow flexible and fair access to spectrum resources. Policymakers and industry stakeholders must collaborate to develop regulations that foster innovation while ensuring fair competition. Future efforts should explore new spectrum management techniques and regulatory policies that balance innovation with equitable access, facilitating effective spectrum utilization.

\textbf{Data Privacy and Sovereignty:} 
Data privacy and sovereignty are critical concerns for 6G and \gls{oran} optimizations. Ensuring compliance with regional regulations while optimizing \gls{ai}-driven networks is a major challenge. Future work should focus on developing privacy-preserving \gls{ai} techniques and data management frameworks that adhere to local laws and regulations, ensuring that data privacy and sovereignty are respected in the evolving digital landscape. This will require collaborative efforts between technology developers, regulators, and policymakers to create frameworks that balance innovation with privacy and legal compliance.

Looking forward, the integration of \gls{oran} and 6G offers tremendous potential for advancing wireless communication. As \gls{ai}-driven networks evolve, they will become increasingly intelligent, optimizing operations, predicting maintenance needs, and enhancing security in real-time. The exploration of the \gls{thz} spectrum and the development of quantum communications will further complement these advancements, creating a robust infrastructure capable of handling future demands. Innovations in edge and fog computing, network slicing, and global collaboration will also be crucial in shaping the future of \gls{oran} and 6G, ensuring that these technologies are not only advanced but also inclusive, secure, and sustainable. By addressing the challenges and pursuing these future directions, the integration of \gls{oran} and 6G can drive the next generation of wireless networks, fostering a more connected, efficient, and adaptive digital ecosystem.
\section{Conclusions}
\label{sec:conclusion}
This paper thoroughly examined the complex interaction between \gls{oran} and 6G, focusing on how they can work together seamlessly. We analyzed the principles of openness, flexibility, intelligence, and cost-efficiency that define \gls{oran}, revealing its potential to enable customization, cost savings, and innovation within 6G networks. We also addressed the challenges posed by 6G’s requirements for extreme performance, non-terrestrial integration, and robust security. By leveraging \gls{oran}’s capabilities, the path to achieving 6G’s ambitious functional promises can be accelerated, laying a solid foundation for future advancements. Furthermore, the paper presented a novel learning-based spectrum-sharing solution, showcasing how \gls{oran} aligns with 6G’s goals of spectrum optimization, underlining its critical role in shaping the future networks. 
\section*{Acknowledgement}
This work is partially funded by the Department of Homeland Security’s Science and Technology Directorate (S\&T).

\section*{Biography}
\label{sec:bio}
\small
\textbf{Sneihil Gopal} received her Ph.D. degree in Electronics and Communications Engineering in $2021$ from IIIT-Delhi, India. She is currently working as a Postdoctoral Researcher at the Department of Physics, Georgetown University, and as PREP Associate in the Wireless Networks Division, NIST. Her research interests are in the general area of communications and networking, dynamic spectrum sharing, wireless network optimization, applications of game theory in wireless networks, and machine learning.

\textbf{Aziz Kord} with many years of experience in the technical aspects of wireless telecommunications and has been deeply involved in the research and development of hardware and software. He has worked with prestigious organizations such as NIST, Motorola, Nokia, and T-Mobile. He has also made significant contributions to multiple industries and has been honored with several accolades, including the Gold, Bronze, and the prestigious Allen V. Astin Measurement Science Awards. His primary interest lies in all aspects of wireless technologies, especially in the RF area.

\textbf{Richard Rouil} received his Ph.D. degree in computer science in 2009 from Telecom Bretagne, France, that focused on mobility in heterogeneous networks. He is currently the Division Chief of the Wireless Networks Division at NIST. His research focuses on the performance evaluation of wireless technologies, such as \gls{lte} and \gls{nr} to support the development, analysis, and deployment of networks used by public safety. His main interests include protocol modeling and simulation of communication networks.
\begin{spacing}{}
    \bibliographystyle{IEEEtran}
    \bibliography{references}

\begin{thebibliography}{10}
\providecommand{\url}[1]{#1}
\csname url@samestyle\endcsname
\providecommand{\newblock}{\relax}
\providecommand{\bibinfo}[2]{#2}
\providecommand{\BIBentrySTDinterwordspacing}{\spaceskip=0pt\relax}
\providecommand{\BIBentryALTinterwordstretchfactor}{4}
\providecommand{\BIBentryALTinterwordspacing}{\spaceskip=\fontdimen2\font plus
\BIBentryALTinterwordstretchfactor\fontdimen3\font minus
  \fontdimen4\font\relax}
\providecommand{\BIBforeignlanguage}[2]{{%
\expandafter\ifx\csname l@#1\endcsname\relax
\typeout{** WARNING: IEEEtran.bst: No hyphenation pattern has been}%
\typeout{** loaded for the language `#1'. Using the pattern for}%
\typeout{** the default language instead.}%
\else
\language=\csname l@#1\endcsname
\fi
#2}}
\providecommand{\BIBdecl}{\relax}
\BIBdecl

\bibitem{ref:ORAN_polese_2023}
M.~Polese, L.~Bonati, S.~D’oro, S.~Basagni, and T.~Melodia, ``{Understanding
  O-RAN: Architecture, interfaces, algorithms, security, and research
  challenges},'' \emph{IEEE Communications Surveys \& Tutorials}, 2023.

\bibitem{ref:6G_IMT2030}
I.~Recommendation, ``{Framework and overall objectives of the future
  development of IMT for 2030 and beyond},'' \emph{International
  Telecommunication Union (ITU) Recommendation (ITU-R)}, 2023.

\bibitem{ref:6G_de_2021_survey}
C.~De~Alwis, A.~Kalla, Q.-V. Pham, P.~Kumar, K.~Dev, W.-J. Hwang, and
  M.~Liyanage, ``{Survey on 6G frontiers: Trends, applications, requirements,
  technologies and future research},'' \emph{IEEE Open Journal of the
  Communications Society}, vol.~2, pp. 836--886, 2021.

\bibitem{ref:oran_tr_wg1}
\vspace{.01mm}{O-RAN} Alliance, ``{O}-{RAN} {W}orking {G}roup 1 {U}se {C}ases
  {A}nalysis {R}eport,'' \emph{O-RAN.WG1.Use-Cases-Analysis-Report-R003}, vol.
  10.00, 2023.

\bibitem{ref:oran_TIP}
``{T}elecom {I}nfra {P}roject,'' \url{https://telecominfraproject.com/},
  accessed 31-08-2024.

\bibitem{ref:oran_arnaz2022toward}
A.~Arnaz, J.~Lipman, M.~Abolhasan, and M.~Hiltunen, ``Toward integrating
  intelligence and programmability in open radio access networks: A
  comprehensive survey,'' \emph{Ieee Access}, vol.~10, pp. 67\,747--67\,770,
  2022.

\bibitem{ref:6G_saad_2019_vision}
W.~Saad, M.~Bennis, and M.~Chen, ``{A vision of 6G wireless systems:
  Applications, trends, technologies, and open research problems},'' \emph{IEEE
  network}, vol.~34, no.~3, pp. 134--142, 2019.

\bibitem{ref:6G_Keysight}
``{Next Generation Wireless: A Guide to the Fundamentals of 6G},''
  \emph{Keysight}, 2024.

\bibitem{ref:ORAN_6G}
V.~Dixit, J.~Plachy, K.~Sun, A.~Ikami, E.~Obiodu, and K.~Lee, ``{O-RAN next
  Generation Research Group (nGRG) Research Report: O-RAN Towards 6G},'' vol.
  RR-2023-01, 2023.

\bibitem{ref:oran_gopal_2024_adapshare}
\BIBentryALTinterwordspacing
{S.Gopal, D. Griffith, R.A. Rouil and C. Liu}, ``{AdapShare: An RL-Based
  Dynamic Spectrum Sharing Solution for O-RAN},'' 2024. [Online]. Available:
  \url{https://arxiv.org/abs/2408.16842}
\BIBentrySTDinterwordspacing

\bibitem{ref:oran_gopal_2024prosas}
S.~Gopal, D.~Griffith, R.~A. Rouil, and C.~Liu, ``{ProSAS: An O-RAN Approach to
  Spectrum Sharing Between NR and LTE},'' in \emph{ICC 2024 - IEEE
  International Conference on Communications}, 2024, pp. 360--366.

\bibitem{ref:rl_sutton2018reinforcement}
R.~S. Sutton and A.~G. Barto, \emph{Reinforcement learning: An
  introduction}.\hskip 1em plus 0.5em minus 0.4em\relax MIT press, 2018.

\bibitem{ref:datset_adapshare}
S.~Gopal, D.~Griffith, R.~Rouil, and C.~Liu, ``{AdapShare: An RL-Based Dynamic
  Spectrum Sharing Solution for O-RAN,},''
  \url{https://doi.org/10.18434/mds2-3613}, accessed 2024-11-06.

\bibitem{ref:owl_bui_2016}
N.~Bui and J.~Widmer, ``{OWL: A reliable online watcher for LTE control channel
  measurements},'' \emph{Proceedings of the 5th Workshop on All Things
  Cellular: Operations, Applications and Challenges}, pp. 25--30, 2016.

\bibitem{ref:timegan}
J.~Yoon, D.~Jarrett, and M.~Van~der Schaar, ``{Time-series Generative
  Adversarial Networks},'' \emph{{Advances in Neural Information Processing
  Systems}}, vol.~32, 2019.

\end{thebibliography}
\end{spacing}
\end{document}